\documentclass[twocolumn,aps,showpacs,prb]{revtex4-1}

\usepackage{amsmath,amssymb,graphics,epsfig,epstopdf,color,multirow,array,verbatim,ulem,braket,tabularx}
\usepackage[colorlinks,linkcolor=blue,citecolor=blue,urlcolor=blue]{hyperref}

\begin{document}

\title{The melilite-type compound (Sr$_{1-x}$,$A_x$)$_2$MnGe$_2$S$_6$O ($A$=K, La) being a room temperature ferromagnetic semiconductor}

\author{Huan-Cheng Yang}
\author{Ben-Chao Gong}
\author{Kai Liu}
\email{kliu@ruc.edu.cn}
\author{Zhong-Yi Lu}
\email{zlu@ruc.edu.cn}
\affiliation{Department of Physics and Beijing Key Laboratory of Opto-electronic Functional Materials $\&$ Micro-nano Devices, Renmin University of China, Beijing 100872, China}

\begin{abstract}
The seeking of room temperature ferromagnetic semiconductors, which take advantages of both the charge and spin degrees of freedom of electrons to realize a variety of functionalities in devices integrated with electronic, optical, and magnetic storage properties, has been a long-term goal of scientists and engineers. Here, by using the spin-polarized density functional theory calculations, we predict a new series of high temperature ferromagnetic semiconductors based on the melilite-type oxysulfide Sr$_2$MnGe$_2$S$_6$O through hole (K) and electron (La) doping. Due to the lack of strong antiferromagnetic superexchange between Mn ions, the weak antiferromagnetic order in the parent compound Sr$_2$MnGe$_2$S$_6$O can be suppressed easily by charge doping with either $p$-type or $n$-type carriers, giving rise to the expected ferromagnetic order. At a doping concentration of 25\%, both the hole-doped and electron-doped compounds can achieve a Curie temperature ($T_\text{c}$) above 300 K. The underlying mechanism is analyzed. Our study provides an effective approach for exploring new types of high temperature ferromagnetic semiconductors.
\end{abstract}

\pacs{75.50.Pp, 71.20.Nr, 71.15.Mb}

\date{\today} \maketitle


\section{Introduction}

Magnetic semiconductors have been a research focus for a long time due to the expectation of simultaneously utilizing charge and spin degrees of freedom of electrons to achieve entirely new functionalities~\cite{Science_Ohno, NatureMaterials_I, RevModPhys_I, RevModPhys_II}. Diluted magnetic semiconductors (DMSs) are a representative family of magnetic semiconductors, which are nonmagnetic semiconductors doped with magnetic dopants. As model systems of DMSs, Mn-doped A$_{1-x}^\text{II}$Mn$_{x}$B$^\text{VI}$ alloys~\cite{JAP_I} and A$_{1-x}^\text{III}$Mn$_{x}$B$^\text{V}$ compounds~\cite{Handbook_I}, which were inspired by the motivation of introducing local moments into well-understood nonmagnetic semiconductors to make them ferromagnetic (FM)~\cite{PSSB_I, PSSB_II}, have attracted extensive attentions. In the Mn-doped A$_{1-x}^\text{II}$Mn$_{x}$B$^\text{VI}$ alloys, the valence of the Mn$^{2+}$ ions matches that of the cations, which allows them to be easily prepared in bulk form. Nevertheless, the early attempt at charge doping, either $p$-type or $n$-type, was difficult and the antiferromagnetic (AFM) superexchange dominates the magnetism of the A$_{1-x}^\text{II}$Mn$_{x}$B$^\text{VI}$ alloys, until the emergence of ferromagnetism at very low temperatures owing to the latter progress in charge doping~\cite{PRL_79.511}. On the other hand, the major obstacle in making III-V semiconductors magnetic is the low equilibrium solubility of Mn in the compounds~\cite{Science_Ohno, Handbook_I}. The successful application of non-equilibrium epitaxial growth technique on the III-V based DMSs allows the solubility of Mn in A$_{1-x}^\text{III}$Mn$_{x}$B$^\text{V}$ films to exceed its equilibrium limited solubility. Subsequently, hole-induced ferromagnetic order in $p$-type (In,Mn)As was reported by Ohno $et$ $al$.~\cite{PRL_68.2664}. And the most widely studied (Ga,Mn)As has achieved a Curie temperature $T_\text{c}$ of 200 K~\cite{Nano_I}. However, the great challenge for the ultimate goal to obtain ferromagnetic DMSs working at room temperature still stands.

Recently, Mn doped I-II-V compound Li(Zn, Mn)As~\cite{NatureCommunications_I}, II-II-V compound (Ba,K)(Zn,Mn)$_2$As$_2$~\cite{NatureCommunications_II}, and III-VI-II-V compound (La,Ba)O(Zn,Mn)As~\cite{PRB_Ning}, which are respectively isostructural to the related iron-based superconductors LiFeAs~\cite{SSC_I}, BaFe$_{2}$As$_{2}$~\cite{PRL_101.107006}, and LaOFeAs~\cite{JACS}, were reported as new types of DMS materials. In these compounds, the spin doping and charge doping are decoupled and both can reach their respective appropriate concentrations. Meanwhile, they can be prepared as bulk samples effectively~\cite{NatureCommunications_I, NatureCommunications_II, PRB_Ning}. Noteworthily, the $T_\text{c}$ of (Ba$_{0.7}$K$_{0.3}$)(Zn$_{0.85}$Mn$_{0.15}$)$_2$As$_2$ even reaches ~230 K~\cite{CSB}, higher than the record value achieved in (Ga,Mn)As~\cite{Nano_I}. However, for all these compounds together with the A$_{1-x}^\text{II}$Mn$_x$B$^\text{VI}$ alloys and the A$_{1-x}^\text{III}$Mn$_{x}$B$^\text{V}$ compounds, the cation sites (Zn or Ga) for spin doping locate in the tetrahedral crystal field of anions (such as As) and each cation site can connect with its nearest- and sometimes next-nearest- neighbor cation sites by bridging anions directly [Figs.~\ref{fig:1}(c) and ~\ref{fig:1}(d)]. Thus, the doped Mn$^{2+}$ ions, which replace the cations, may form Mn-Mn pairs bridged directly by anions (indeed true in the experiments~\cite{NatureCommunications_I, NatureCommunications_II, PRB_Ning}) so as to be antiferromagnetically coupled through superexchange interactions favorably in energy~\cite{SSC_II, SSP_I, PRB_V, EPJB_I}. This induces the reduction of the net average magnetization for the Mn$^{2+}$ ions and is harmful to the improvement of sample quality. On the other hand, for all these compounds, the FM order appears only when the charge doping is introduced, demonstrating that the  ferromagnetism is induced by itinerant carriers~\cite{PRL_79.511, PRL_68.2664, NatureCommunications_I, NatureCommunications_II, PRB_Ning}. Based on the above facts, searching for such a material, in which the nearest-neighbor AFM coupling is avoided and the effective charge doping can be achieved, may serve as an effective approach for exploring more feasible magnetic semiconductors~\cite{Yang}.

Recently, the melilite-type oxide Sr$_2$MnGe$_2$O$_7$~\cite{InorgChem_I} and oxysulfide Sr$_2$MnGe$_2$S$_6$O~\cite{InorgChem_II} were successfully synthesized in experiments and reported as weakly AFM insulators with respective magnetic transition temperatures being only 4.4 K and 15.5 K. Structurally, in both compounds, the MnO$_4$ (MnS$_4$) tetrahedra are separated by the GeO$_4$ (GeS$_3$O) tetrahedra [Figs.~\ref{fig:1}(a) and ~\ref{fig:1}(b)] and the Mn atoms are distributed uniformly at a moderate separation of 5.86-6.73 {\AA}~\cite{InorgChem_I, InorgChem_II}. This provides a great opportunity to block the possible nearest-neighbor AFM superexchange. In addition, the sandwiched Sr atoms [Fig.~\ref{fig:1}(b)] act as charge reservoir layers similar to the Ba atoms in BaZn$_2$As$_2$ [Fig.~\ref{fig:1}(c)], which is promising for a feasible charge doping. Inspired by the rising of FM order in all previous DMSs via introducing itinerant carries~\cite{PRL_79.511, PRL_68.2664, NatureCommunications_I, NatureCommunications_II, PRB_Ning}, we infer that the charge doping may readily suppress the weak AFM order and further induce the ferromagnetic order in these melilite-type compounds.

In this work, we have systematically studied the magnetic properties and electronic structures of the charge-doped oxysulfide Sr$_2$MnGe$_2$S$_6$O as a model system for the melilite-type compounds by using density functional theory (DFT) calculations. We find that the charge doping, either $p$-type or $n$-type, can suppress the AFM order in the parent compound and then raise the FM order in the doped compounds (Sr$_{1-x}$,K$_x$)$_2$MnGe$_2$S$_6$O and (Sr$_{1-x}$,La$_x$)$_2$MnGe$_2$S$_6$O. With a carrier concentration of 25\%, typically the one in the previous DMS material (Ba$_{1-x}$K$_x$)(Zn$_{1-y}$Mn$_y$)$_2$As$_2$~\cite{NatureCommunications_II}, the (Sr$_{0.75}$,$A_{0.25}$)$_2$MnGe$_2$S$_6$O ($A$=K, La) compound can achieve a Curie temperature $T_\text{c}$ above 300 K. The rigid shifts of the band structures after doping show the semiconductor characteristics. Our theoretical prediction on this new type of room temperature ferromagnetic semiconductors calls for experimental validation.

\section{Computational details}
\label{sec:Method}
\begin{figure}[!t]
\centering
\includegraphics[width=0.99\columnwidth]{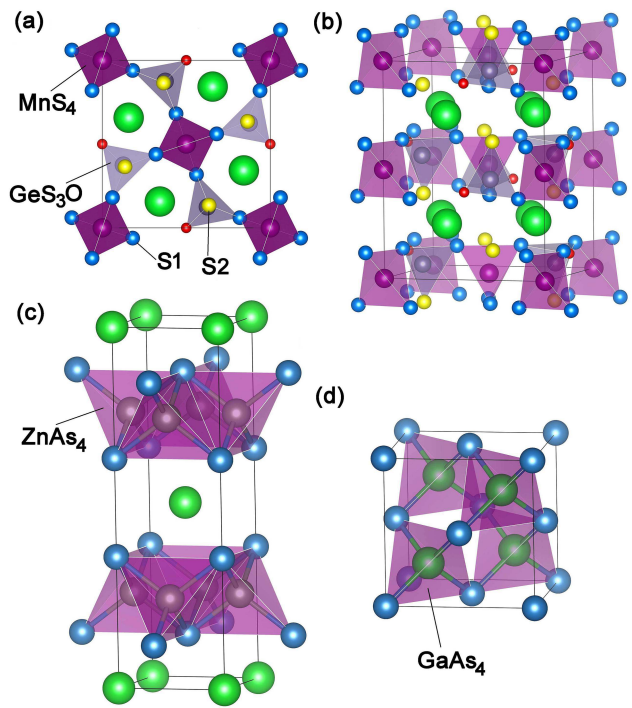}
\caption{\small(Color online) (a) Top view and (b) Side view of the crystal structure for Sr$_2$MnGe$_2$S$_6$O, in which the MnS$_4$ tetrahedra are separated by the GeS$_3$O tetrahedra. Side views of the crystal structures for (c) BaZn$_2$As$_2$ and (d) GaAs, in which the ZnAs$_4$ (GaAs$_4$) tetrahedra connect directly with each other.}
\label{fig:1}
\end{figure}

The spin-polarized density functional theory (DFT) calculations were carried out by using the projector augmented wave (PAW) method~\cite{PAW_I, PAW_II} as implemented in the Vienna Ab initio Simulation Package~\cite{VASP_I, VASP_II, VASP_III}. The generalized gradient approximation (GGA) of Perdew-Burke-Ernzerhof (PBE) was employed for the exchange-correlation functional~\cite{PBE}. The kinetic energy cutoff of the plane-wave basis was set to be 520 eV. The space group of Sr$_2$MnGe$_2$S$_6$O is P$\overline{4}$2$_1$m~\cite{InorgChem_II}. In order to describe different magnetic configurations, we chose a supercell containing 48 atoms by doubling the tetragonal cell along $c$-direction as shown in Fig.~\ref{fig:1}(b). To simulate the hole-type or electron-type doping, a part of Sr atoms were substituted by K atoms or La atoms, respectively. The substitution of one Sr atom in the supercell corresponds to a charge-doping concentration of 12.5\%, and that of two Sr atoms in different Sr layers to 25\%. The lattice constants were fixed at the experimental values ($a=b=9.5206$ {\AA} and $2c=2\times6.2002=12.4004$ {\AA})~\cite{InorgChem_II}. The Brillouin zone was sampled with a $7\times7\times6$ Monkhorst-Pack $k$-point mesh. All internal atomic positions were optimized until the forces on atoms reached a convergence criterion 0.01 eV/{\AA}. In addition, we double checked the results by employing the virtual crystal approximation (VCA) to simulate various charge doping concentrations in (Sr$_{1-x}$,$A_x$)$_2$MnGe$_2$S$_6$O ($A$=K, La). The correlation effect in the 3$d$ orbitals of Mn atoms was checked in the GGA+U framework by using the simplified rotationally invariant version of Dudarev \textit{et al.} (effective U)~\cite{DFTU-II}.

\section{Results and analysis}
\label{sec:Results}

In the melilite-type oxysulfide Sr$_2$MnGe$_2$S$_6$O, the different layers formed by corner sharing MnS$_4$ and GeS$_3$O tetrahedra are separated by Sr atoms, while the Mn atoms in the same layer form a square lattice [Fig.~\ref{fig:1}(a)]. Previous neutron scattering experiments determined that the magnetic ground state of Sr$_2$MnGe$_2$S$_6$O is in an intra-layer checkerboard AFM N\'eel order with a very weak inter-layer AFM coupling~\cite{InorgChem_II}. For the doped compound (Sr$_{1-x}$,$A_x$)$_2$MnGe$_2$S$_6$O ($A$=K, La), we thus mainly considered two typical magnetic orders: the same checkerboard AFM N\'eel order as the one in its parent compound as well as an FM order with both intra-layer and inter-layer ferromagnetic ordering. Moreover, our calculations demonstrate that the inter-layer magnetic coupling in (Sr$_{1-x}$,$A_x$)$_2$MnGe$_2$S$_6$O is also very weak compared with the intra-layer one, suggesting its two dimensional magnetic characteristics.

\begin{figure}[!t]
\centering
\includegraphics[width=0.99\columnwidth]{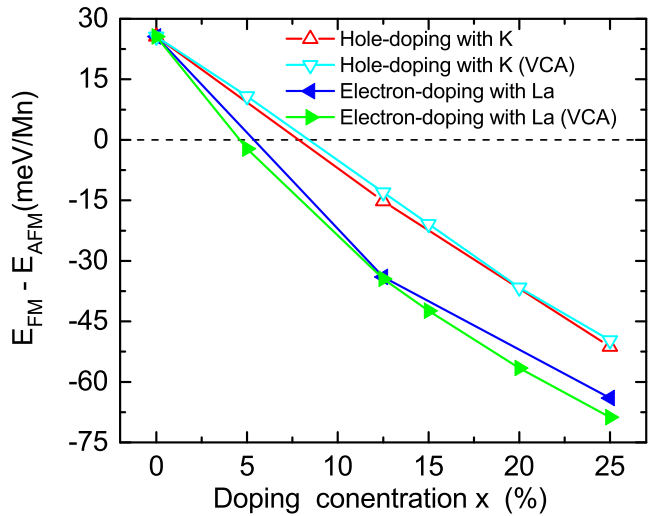}
\caption{\small(Color online)  Energy differences between the ferromagnetic (FM) and antiferromagnetic (AFM) orders for (Sr$_{1-x}$,$A_x$)$_2$MnGe$_2$S$_6$O ($A$=K, La) with different hole- and electron- doping concentrations.}
\label{fig:2}
\end{figure}

Figure~\ref{fig:2} shows the energy difference between the FM and AFM orders of the compound (Sr$_{1-x}$,$A_x$)$_2$MnGe$_2$S$_6$O as a function of charge doping concentration $x$. At zero charge doping, the magnetic ground state of Sr$_2$MnGe$_2$S$_6$O is in the AFM order, which is in accordance with the neutron scattering measurement~\cite{InorgChem_II}. With one K (La) atom substituting one Sr atom, equivalent to introducing 12.5\% holes (electrons) into the compound, the FM order is energetically lower by an energy of about 15 (34) meV/Mn than the AFM order. When the doping concentration increases to 25\%, namely two K (La) atoms substituting two Sr atoms, the magnitudes of the energy differences between the FM and AFM orders further expand to about 51 and 64 meV/Mn for (Sr$_{0.75}$,K$_{0.25}$)$_2$MnGe$_2$S$_6$O and (Sr$_{0.75}$,La$_{0.25}$)$_2$MnGe$_2$S$_6$, respectively. Thus, both the hole-type and electron-type carriers can suppress the AFM order in (Sr$_{1-x}$,$A_x$)$_2$MnGe$_2$S$_6$O and favor the FM order. It is worth noting that our VCA results agree well with the ones calculated by real atomic substitution. In addition, the data from the VCA calculations for other doping concentrations almost locate on the lines outlined by the data from the direct atomic substitution calculations (Fig.~\ref{fig:2}), which demonstrates that the VCA calculations can describe (Sr$_{1-x}$,$A_x$)$_2$MnGe$_2$S$_6$O reasonably. Following the trend of the VCA results, the FM order will be energetically further lower than the AFM order at higher doping concentrations.

\begin{figure}[!t]
\centering
\includegraphics[width=0.99\columnwidth]{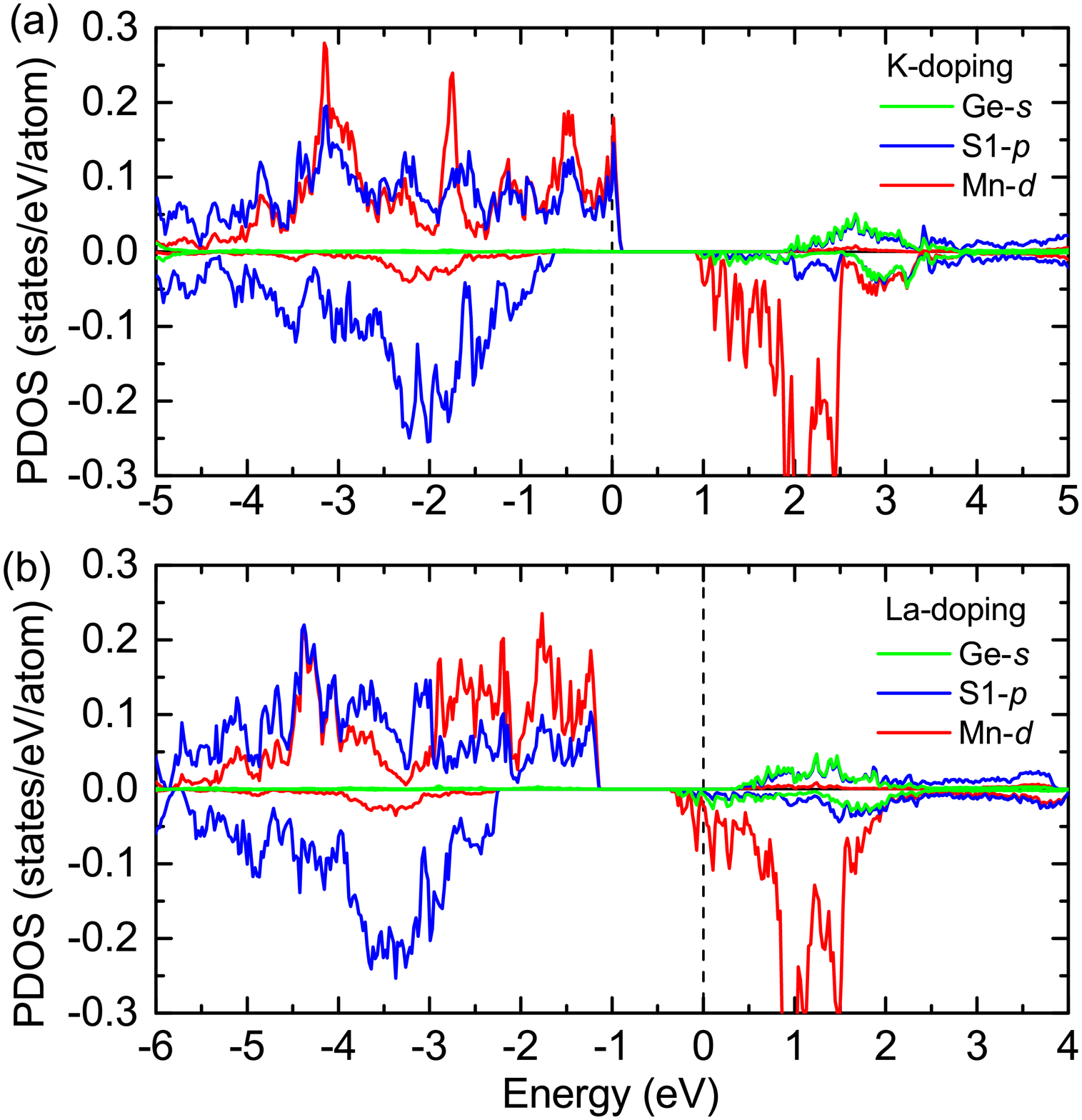}
\caption{\small(Color online) Partial density of states (PDOS) for (a) (Sr$_{0.875}$,K$_{0.125}$)$_2$MnGe$_2$S$_6$O and (b) (Sr$_{0.875}$,La$_{0.125}$)$_2$MnGe$_2$S$_6$O. The S1 atom is labeled in Fig.~\ref{fig:1}(a). The up and down parts of each panel correspond to the spin-up and spin-down channels, respectively. The Fermi energy sets to zero.}
\label{fig:3}
\end{figure}

\begin{figure*}[!t]
\centering
\includegraphics[width=0.95\textwidth]{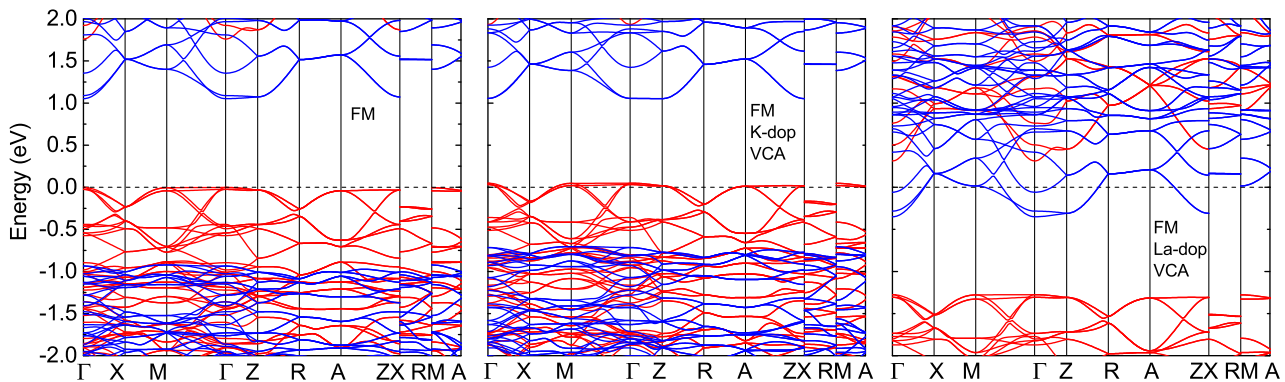}
\caption{\small(Color online) Electronic band structures for Sr$_2$MnGe$_2$S$_6$O, (Sr$_{0.875}$,K$_{0.125}$)$_2$MnGe$_2$S$_6$O, and (Sr$_{0.875}$,La$_{0.125}$)$_2$MnGe$_2$S$_6$O, respectively. The Sr$_2$MnGe$_2$S$_6$O is set to be in the FM order. The red lines denote the spin-up channels, while the blue lines represent the spin-down ones.}
\label{fig:4}
\end{figure*}

In order to well understand the doped compound (Sr$_{1-x}$,$A_x$)$_2$MnGe$_2$S$_6$O, we further inspected the electronic structures of (Sr$_{0.875}$,$A_{0.125}$)$_2$MnGe$_2$S$_6$O for illustration. The partial density of states (PDOS) for those atomic species, whose orbitals may have main contributions around the Fermi level, are shown in Figure~\ref{fig:3}. The hole (electron)-type carriers introduced by K (La) doping can be clearly observed. The difference between the K and La doping cases is that besides Mn $d$ orbitals the former brings itinerant holes in S $p$ orbitals while the latter introduces itinerant electrons in Ge $s$ orbitals. The coinciding peaks between the $d$ orbitals of Mn atom and the $p$ orbitals of its neighboring S1 atoms [labeled in Fig.~\ref{fig:1}(a)] extend from -4 eV to the Fermi level for K doping and from -5 eV to -1 eV for La doping, respectively. This indicates the strong $p$-$d$ hybridization. Due to the strong oxidability of O atom and strong reducibility of Sr/K/La atom, their orbital distributions are relatively far away from the Fermi level. Figure \ref{fig:4} shows the electronic band structures of (Sr$_{0.875}$,$A_{0.125}$)$_2$MnGe$_2$S$_6$O, together with the electronic band structure of the undoped compound Sr$_2$MnGe$_2$S$_6$O with the FM order for reference. 
For both the K- and La- doped compounds, the electronic band profiles remain almost unchanged with only rigid shifts of the Fermi level (Fig.~\ref{fig:4}), which is a typical feature for doped semiconductors.

The GGA+U calculations were also carried out to check the correlation effect in the 3$d$ orbitals of Mn atoms by using the simplified rotationally invariant version of Dudarev \textit{et al.} (effective U)~\cite{DFTU-II}. The values of the effective U we used are 2, 4, and 6 eV, respectively, while 6 eV is large enough for describing the correlation effect in divalent Mn sulfides~\cite{RevModPhys_III}. We find that the consideration of the Hubbard U does not change the fact that the FM order appears after both the hole- and electron- doping in Sr$_2$MnGe$_2$S$_6$O. Moreover, the band gaps from the GGA+U calculations for the parent compound Sr$_2$MnGe$_2$S$_6$O with the ground state AFM order are about 1.4, 2.1, 2.3, and 2.4 eV for the U = 0, 2, 4, and 6 eV, respectively. The calculated band gaps are smaller than the experimental value of 3.2 eV~\cite{InorgChem_II}. This ascribes to the well-known fact that the GGA-type DFT calculations underestimate the band gaps for semiconductors and insulators~\cite{gap1, gap2}. The calculations thus show that the calculated band gap for the AFM state of the parent compound is insensitive to the large U parameter, which is consistent with the previous work~\cite{InorgChem_II}. And the similar U dependence also happens in the doped compound (Sr$_{0.875}$,$A_{0.125}$)$_2$MnGe$_2$S$_6$O ($A$=K, La). Thus, the finite band gaps calculated with and without the Hubbard U as well as the rigid band shifts after charge doping further confirm the semiconductor characteristics of the compound (Sr$_{1-x}$,$A_x$)$_2$MnGe$_2$S$_6$O.

The large energy difference between the FM order and the AFM order in the compound (Sr$_{1-x}$,$A_x$)$_2$MnGe$_2$S$_6$O at a high level of charge doping (Fig.~\ref{fig:2}) shows that the FM coupling between the Mn atoms is robust. This may result from the effect of the strong $p$-$d$ hybridization between the $d$ orbitals of Mn atoms and the $p$ orbitals of its neighboring S1 atoms (Fig.~\ref{fig:3}) in combination with the transmission effect of $p$-$d$ exchange interactions by itinerant carriers, which eventually results in the long-range effective FM coupling among the moderately separated Mn spins. In the compound (Sr$_{1-x}$,$A_x$)$_2$MnGe$_2$S$_6$O, the (Sr$_{1-x}$,$A_x$) layers act as charge reservoir and structurally they isolate different Mn-Ge layers [Fig.~\ref{fig:1}(b)]. The calculations show that overall the spin-parallel is just 1.0-1.5 meV/Mn lower in energy than the spin-antiparallel between the Mn-Ge layers, which indicates a weak interlayer magnetic coupling between different Mn-Ge layers. Thus the compound (Sr$_{1-x}$,$A_x$)$_2$MnGe$_2$S$_6$O is a quasi-two-dimensional ferromagnet.

To quantify the magnetism in the compound (Sr$_{1-x}$,$A_x$)$_2$MnGe$_2$S$_6$O, we map the energy difference between the FM order and the checkerboard AFM N\'eel order onto a two-dimensional square Heisenberg model with the nearest-neighbor coupling J$_1$ as follows,
\begin{equation}
H = -J_{1}\sum_{<ij>}\textbf{e}_{i}\cdot{\mathbf{e}_{j}} .
\end{equation}
Here, \textbf{e}$_{i}$ is the unit vector in the direction of the spin \textbf{S}$_{i}$ on site $i$ with the moment $M$ (\textbf{e}$_{i}$ = \textbf{S}$_{i}$/$M$) and $J_{1}$ is the exchange integral parameter including the contribution of the magnitude of the moment $M$. The relationship between $J_1$ and the energy difference per Mn between the AFM order and the FM order, denoted as $\triangle$$E$ = $E_\text{AFM}$-$E_\text{FM}$, is $J_1$ = $\triangle$$E$/4. In all the calculations, the magnetic moment on each Mn atom is about 4 $\mu_{B}$ (spin S=2). For such large magnetic moments, the possible magnetic quantum fluctuations can be neglected. Thus, we can estimate the Curie temperature $T_\text{c}$ through the analytical solution of a two-dimensional square lattice Ising model denoted as $k_\text{B}$$T_\text{c}$ = 2$J_1$/(ln(1+$\sqrt{2}$)~\cite{ISING}. At a doping concentration of 25\%, both the K- and La- doped (Sr$_{0.75}$,$A_{0.25}$)$_2$MnGe$_2$S$_6$O can achieve a $T_\text{c}$ above 300 K, namely 327 K and 453 K (see Table 1) respectively. This prediction for this new type of \textit{room temperature} ferromagnetic semiconductors waits for experimental validation.

\begin{table*}[!t]
\caption{\small\label{tab:I}The lattice constants for melilite-type compounds, as well as the estimated Curie temperature $T_\text{c}$ at 25\% hole-doping (K) and electron-doping (La). Here, for the charge doping, the VCA method was used without further cell relaxation.}
\renewcommand\arraystretch{1.4}
\begin{center}
\begin{tabular*}{0.75\textwidth}{@{\extracolsep{\fill}}ccccc}
\hline\hline
  \multirow{2}{*}{Compounds} &  \multicolumn{2}{c}{Lattice constants ({\AA})}& \multicolumn{2}{c}{$T_\text{c}$ (K)} \\
  \cline{2-3} \cline{4-5}
  & $a=b$&  2$c$&  25\%-K&  25\%-La \\
\hline
  $^\textrm{a}$Sr$_2$MnGe$_2$S$_6$O&  9.5206&  12.4004&  327&  453  \\
  $^\textrm{b}$Sr$_2$MnGe$_2$S$_7$&   9.9403&  12.7269&  153&  365  \\
  $^\textrm{b}$Sr$_2$MnGe$_2$Se$_7$&  10.4461&  13.2996&  89&  197  \\
  $^\textrm{b}$Ba$_2$MnGe$_2$S$_6$O&  9.8339&  12.9877&  268&  384  \\
  $^\textrm{b}$Ba$_2$MnGe$_2$S$_7$&   10.1739& 13.1614&   83&  311  \\
  $^\textrm{b}$Ba$_2$MnGe$_2$Se$_7$&  10.6829& 13.6805&   13&  188  \\
  $^\textrm{b}$Sr$_2$MnSi$_2$S$_6$O&  9.4796&  12.4138&  345&  301  \\
  $^\textrm{b}$Sr$_2$FeGe$_2$S$_7$&   9.8793&  12.7119&   55&  106  \\
  $^\textrm{b}$Sr$_2$FeGe$_2$Se$_7$&  10.3754& 13.2164&   46&  138  \\
\hline\hline
\end{tabular*}
\small
\begin{flushleft}
$^\textrm{a}$ From the experimental data\cite{InorgChem_II}. \\
$^\textrm{b}$ From fully optimized calculations.
\end{flushleft}
\end{center}
\end{table*}

\section{Discussion and Summary}
\label{sec:discussion}

The most prominent feature in this new melilite-type ferromagnetic semiconductors (Sr$_{1-x}$,$A_x$)$_2$MnGe$_2$S$_6$O is that the Mn atoms distribute uniformly with a moderate separation. Accordingly there are several advantages. First, the moderate separation between magnetic Mn ions avoids the strong AFM superexchange and results in a weak AFM order in the undoped parent compound, which can be suppressed easily, and then the FM order is readily induced after charge doping (Fig.~\ref{fig:2}). Second, the not-too-far Mn-Mn separation is helpful for the effective FM coupling between the Mn spins. Otherwise, the FM interaction will become negligible at a quite large Mn-Mn separation. Third, compared with the previously reported DMSs such as Li(Zn,Mn)As, (Ba,K)(Zn,Mn)$_2$As$_2$, and (La,Ba)O(Zn,Mn)As, in which the antiferromagnetically coupled nearest-neighbor Mn-Mn pairs substantially reduce the average net magnetization, all Mn ions in (Sr$_{1-x}$,$A_x$)$_2$MnGe$_2$S$_6$O can contribute to the ferromagnetism. Forth, the inherent uniformly distributed magnetic Mn ions in (Sr$_{1-x}$,$A_x$)$_2$MnGe$_2$S$_6$O are beneficial for the improvement of sample homogeneity and make only the charge doping need to be considered. Last but not least, both the $p$-type doping and the $n$-type doping are available, which is promising for potential device applications. Thus our study on (Sr$_{1-x}$,$A_x$)$_2$MnGe$_2$S$_6$O serves as an effective approach for exploring new FM semiconductors.

Inspired by the study on (Sr$_{1-x}$,$A_x$)$_2$MnGe$_2$S$_6$O, we further performed calculations for a series of melilite-type compounds at a typical concentration (25\%) of charge doping. The calculated results are reported in Table~\ref{tab:I}. Besides the doped Sr$_2$MnGe$_2$S$_6$O, several other melilite-type compounds, if properly doped, are very promising to show Curie temperatures above room temperature.

In summary, we have investigated the magnetic properties and electronic structures of the hole- and electron- doped melilite-type oxysulfide compound (Sr$_{1-x}$,$A_x$)$_2$MnGe$_2$S$_6$O ($A$=K, La) as a model system by using density functional theory calculations. Due to the inherent crystal structure with relatively large Mn-Mn separations, the strong AFM superexchange between the nearest-neighbor Mn atoms which prevails in other magnetic semiconductors can be naturally avoided. By introducing itinerant carriers, either $p$-type or $n$-type, into the parent compound Sr$_2$MnGe$_2$S$_6$O, the original weak AFM order is easily suppressed and then the FM order is established in (Sr$_{1-x}$,$A_x$)$_2$MnGe$_2$S$_6$O ($A$=K, La). At a doping concentration of $x$ = 25\%, both the hole- and electron- doped compounds can achieve a Curie temperature $T_\text{c}$ above room temperature (300 K). The partial density of states as well as the band structures with rigid shifts of the Fermi level shows the semiconductor characteristics of (Sr$_{1-x}$,$A_x$)$_2$MnGe$_2$S$_6$O. Our study on these melilite-type compounds suggests an effective approach for exploring new types of FM semiconductors from the intrinsic weakly coupled AFM insulators with relatively large M-M (magnetic ions) separation.

\begin{acknowledgments}
This work was supported by the National Key R\&D Program of China (Grant No. 2017YFA0302903), the National Natural Science Foundation of China (Grants No. 11774422 and No. 11774424). Computational resources were provided by the Physical Laboratory of High Performance Computing at Renmin University of China.

Huan-Cheng Yang and Ben-Chao Gong contributed equally to this work.
\end{acknowledgments}

\end{document}